\documentclass[12pt]{article}

\usepackage{amsmath}
\usepackage{amssymb}
\usepackage{bbold}
\usepackage{color}
\usepackage{graphicx}
\usepackage{amsthm}
\usepackage{multirow}
\usepackage{threeparttable}
\usepackage{authblk}

\newtheorem{theorem}{Theorem}

\theoremstyle{definition}

\theoremstyle{remark}

\title{Ranking nodes in directed networks via continuous-time quantum walks}

\author[1]{Paola Boito}
\author[2]{Roberto Grena}
\affil[1]{Dipartimento di Matematica, Universit\`a di Pisa, Largo B. Pontecorvo 5, 56127 Pisa, Italy}
\affil[2]{C. R. ENEA Casaccia, via Anguillarese 301, 00123 Roma, Italy}
\date{}

\begin{document}

\maketitle

\section*{Abstract}

Four new centrality measures for directed networks based on unitary, continuous-time quantum walks (CTQW) in $n$ dimensions -- where $n$ is the number of nodes  -- are presented, tested and discussed. The main idea behind these methods consists in re-casting the classical HITS and PageRank algorithms as eigenvector problems for symmetric matrices, and using these symmetric matrices as Hamiltonians for CTQWs, in order to obtain a unitary evolution operator. 

The choice of the initial state is also crucial. Two options were tested: a vector with uniform occupation and a vector weighted w.r.t.~in- or out-degrees (for authority and hub centrality, respectively). Two methods are based on a HITS-derived Hamiltonian, and two use  a PageRank-derived Hamiltonian. Centrality scores for the nodes are defined as the average occupation values. 

All the methods have been tested on a set of small, simple graphs in order to spot possible evident drawbacks, and then on a larger number of artificially generated larger-sized graphs, in order to draw a comparison with classical HITS and PageRank. Numerical results show that, despite some pathologies found in three of the methods when analyzing small graphs, all the methods are effective in finding the first and top ten nodes in larger graphs. We comment on the results and offer some insight into the good accordance between classical and quantum approaches.

\section{Introduction}\label{sec:intro}

Centrality measures for networks \cite{centrality1,centrality2} can be loosely defined as measures of importance of nodes. Notions of centrality have received considerable interest in the last decades, especially boosted by the important task of finding efficient algorithms for ranking web pages in search engines. Two classical examples of algorithms for computing centrality scores are HITS \cite{hits} and PageRank \cite{BrinPage}.

In directed networks, nodes can be ranked according to their importance as {\it hubs}  or {\it authorities}. Roughly speaking, a strong hub is a node that points to many important nodes and a strong authority is a node that is pointed to by many important nodes. The notion of ``important nodes'' depends on the method: in HITS, which simultaneously provides hubs and authority scores, good hubs point to good authorities and good authorities are pointed to by good hubs, whereas in PageRank a node has a higher authority centrality score if it is pointed to by strong authorities. 

Algorithms for webpage ranking usually compute authority centrality, but any authority method can be easily converted to a hub method by applying it to the reverse graph obtained by inverting the direction of each link. In matrix terms, if $A$ is the adjacency matrix of the original graph, the reverse graph has adjacency matrix $A^T$. For instance, PageRank is an authority ranking algorithm, but it can be easily converted to hub ranking (Reverse PageRank, see e.g. \cite{Fogaras, Zhirov}) by applying it to $A^T$ instead of $A$. HITS, on the other hand, is fundamentally a power method that alternates between $A$ and $A^T$ and yields both hub and authority rankings.

An interesting feature of PageRank is the fact that it can be seen as a random-walk problem: the Google matrix $G$ built from $A$ is a stochastic matrix, thus representing a random walk, and the ranking score of a node is the asymptotic probability of finding the walker in that node. This is an example of the important role that random walks play in many ranking methods.

The increasing development of research in quantum computation has sparked an interest in finding quantum-formulated ranking algorithms for networks. The relation between node-ranking algorithms and classical random walks suggests that it may be useful to explore the theory of quantum walks on networks, for ranking purposes. On the theory of quantum walks, we mention, among others, the seminal papers \cite{Aharonov1, Aharonov2, Kempeintro, Szegedy} and the review \cite{Venegas}. A good introduction can be found, for instance, in the recent book \cite{Portugal}. Among other features, quantum walks exhibit faster diffusion  w.r.t.~classical random walks, thus allowing for algorithmic speedup \cite{Childs03, Kempe}. 

In contrast to classical random walks, unitary quantum walks typically do not converge to a stationary distribution, because of unitary evolution.
Therefore, time-averaging is expected to be required for the extraction of centrality scores, unless one resorts to the use of mixed quantum-classical evolution \cite{PRsemiclass} or tunable time-dependent Hamiltonian for adiabatic computations \cite{PRadiabatic}. Another remarkable feature is that the average occupation of each node may depend on the initial state, i.e., the prescription of the initial state is part of the ranking method.

Like classical random walks, quantum walks can be defined either in a discrete-time (DTQW) or a continuous-time (CTQW) framework \cite{Childs04, revCTQW}. 
Here we focus on the continuous-time case, which has the merit of taking place in a Hilbert space of dimension equal to the number of nodes, whereas DTQW typically require a dimension of the order of the number of edges.

The formulation of a quantum method for centrality measures based on DTQW or CTQW has been explored by several authors \cite{berry, Paparo, Rossi, Izaac, IzaacPT, Loke, Chawla, QWexperim, QWmultilayer}. \cite{Paparo} proposes a method, called {\em Quantum PageRank}, based on a DTQW that evolves in a Hilbert space of dimension $n^2$, where $n$ is the number of nodes. In CTQW-based methods such as \cite{Izaac, Rossi}, the quantum walk takes place in a Hilbert space of dimension $n$; however, such methods are applicable to undirected graphs only. The authors of \cite{IzaacPT} have circumvented the difficulty of obtaining an evolution operator from a non-symmetric $A$ by forgoing unitarity, and adopting instead non-standard PT-symmetric Hamiltonians.

In a recent work \cite{BoitoGrena}, three unitary-CTQW-based centrality measures for directed networks were introduced and discussed. Unitarity was obtained by defining the CTQW on the associated bipartite graph, following the idea behind \cite{BEK}, hence with a number of degrees of freedom equal to $2n$. One of the methods proved to be especially robust 
and its rankings were very well related to HITS rankings; however, no satisfactory results were obtained when trying to design an algorithm well-related to PageRank. Moreover, the doubling of the dimension of the Hilbert space is a drawback, only partially mitigated by the fact that two of the three proposed methods compute hub- and authority-centrality in a single run.

However, doubling the dimension of the Hilbert space could be unnecessary. Classical HITS and PageRank methods can both be reframed as eigenvector problems on $n \times n$ symmetric matrices. This fact suggests the idea of using such matrices as Hamiltonians to define a unitary $n$-dimensional CTQW on the network. Rankings obtained by such a CTQW can be expected to yield results similar to HITS and PageRank, respectively.

In this paper we explore this possibility, proposing two evolution operators for CTQW in $n$ dimensions, based on a (slightly modified) HITS-derived Hamiltonian and on a PageRank-derived Hamiltonian. As mentioned above, the initial state must be supplied as well, in order to properly define the method. Two choices were studied: a uniformly occupied initial state and an initial state with occupation numbers weighted w.r.t.~ node degrees (in-degrees for authority centrality or out-degrees for hub centrality). On the whole, we propose four new algorithms, combining the two choices for the Hamiltonian and the two initial states.

Section \ref{sec:background} provides the background to understand the problem and recalls the HITS and PageRank algorithms, formulating them as eigenvector problems on symmetric $n \times n$ matrices. It also outlines the main ideas behind quantum centrality algorithms. Our CTQW-based methods are presented in Section \ref{sec:main}. Section \ref{sec:experiments} is devoted to numerical experiments: tests on simple toy models, which allow us to discuss certain features of the methods, are followed by tests on more than 5000 larger graphs of different sizes (from 128 to 1024 nodes). These tests are designed to check whether the obtained rankings are in good accordance with the results of the classical methods they are derived from. Three indicators are used: the probability of finding the same top-ranked node, the number of common nodes in the top-ten ranking, and Kendall's $\tau$ \cite{Ktau}. Sections \ref{sec:discussion} and \ref{sec:conclusions} contain a discussion of the results and concluding remarks.

\section{Background}\label{sec:background}

A {\em (directed) graph} ${\mathcal G}$ is a pair $(V,E)$, where $V$ is the set of nodes, labeled from $1$ to $n$, and $E=\{(i,j)|i,j\in V\}$ is the set of edges. The graph is called {\em undirected} if $(i,j)\in E$ implies $(j,i)\in E$. The graphs we consider are weakly connected, unweighted and contain no loops (i.e., edges from one node to itself) or multiple edges.

The {\em adjacency matrix} associated with ${\mathcal G}$ is the $n\times n$ matrix $A=(A_{ij})$ such that $A_{ij}=1$ if t$(i,j)\in E$, and $A_{ij}=0$ otherwise. Clearly $A$ is symmetric if and only if ${\mathcal G}$ is undirected.

The {\em in-degree} ${\rm deg}_{\rm in}(i)$ of node $i$ is the number of nodes pointing to node $i$, whereas its {\em out-degree} ${\rm deg}_{\rm out}(i)$ is the number of nodes pointed to by node $i$. It holds ${\rm deg}_{\rm in}(i)=({\mathbb{1}}^T A)_i$ and ${\rm deg}_{\rm out}(i)=(A {\mathbb{1}})_i$, where ${\mathbb{1}}\in\mathbb{R}^n$ is the  vector with all entries equal to $1$.

\subsection{Classical centrality measures for directed graphs}\label{subs:classical}

Among the many centrality measures proposed in the literature for directed graphs, we recall two popular ones that inspired the present work. Both of them were originally developed for ranking web pages.

\paragraph{HITS (Hyperlink-Induced Topic Search) \cite{hits}.}

This is an iterative scheme that computes an authority score and a hub score for each node. Each iteration takes the form
\begin{equation}
{ x}^{(k)}=A^T { y}^{(k-1)},\qquad { y}^{(k)}=A { x}^{(k)},\label{eq:hits}
\end{equation}
followed by normalization in $2$-norm, where ${ x}^{(k)}$ and ${ y}^{(k)}$ are the vectors of authority and hub scores, respectively, at the $k$-th step. Authority and hub centralities are defined as the limit values of the corresponding scores for $k\rightarrow\infty$. The initial vector is usually chosen as $\frac{1}{\sqrt{n}}{\mathbb{1}}$.

This process can also be reframed as a power method, yielding the dominant eigenvector of the symmetric matrices $A^T A$ (for authorities) and $A A^T$ (for hubs). 

\paragraph{PageRank \cite{BrinPage, LangvilleMeyer}.} Here we recall the PageRank method for authority scores, as originally conceived for web page ranking. Hub centralities can be obtained via {\em Reverse PageRank} \cite{Fogaras, Chep}, that is, PageRank applied to the ``reversed graph'' with adjacency matrix $A^T$.

In the PageRank algorithm, the adjacency matrix $A$ of the graph is modified to obtain the so-called patched adjacency matrix $\tilde{A}$, which is row-stochastic:
$$
\tilde{A}_{ij}=\left\{\begin{array}{ll}
A_{ij}/{\rm deg_{\rm out}}(i) & {\rm if }\, {\rm deg_{\rm out}}(i)>0,\\
1/n & {\rm otherwise.}
\end{array}\right.
$$
This can be seen as the transition matrix of a random walk: each entry $\tilde{A}_{ij}$ is the probability, for a walker placed at node $i$, to reach node $j$, with the assumption that from each dangling node -- i.e., a node with zero out-degree -- the walker can reach all other nodes with uniform probability. Next, one introduces a teleportation effect by adding a rank-one correction:
\begin{equation}
G=\alpha\tilde{A}+\frac{1-\alpha}{n} {\mathbb 1} {\mathbb 1}^T,\label{eq:googlematrix}
\end{equation} 
where $\alpha\in[0,1]$ is a parameter, usually chosen as $\alpha=0.85$. The Perron-Frobenius theorem ensures that $1$ is the (simple) eigenvalue of maximum modulus for $G^T$. The associated eigenvector $x>0$ such that $G^Tx=x$ yields the PageRank scores, and it can be computed efficiently via the power method.

The Google matrix $G$ as defined above yields authority rankings. For hub rankings, one can define a matrix $G_h$ in the same way, starting from $A^T$ instead of $A$. Note that $G_h \neq G^T$. 

The PageRank method can be rewritten in terms of the symmetric matrix
$$
H_G = (I-G)(I-G)^T.
$$
Indeed, it is easily seen that the problem of finding the dominant eigenvector of $G^T$ is equivalent to the problem of finding a vector in the null space of $H_G$. Obviously the dominant eigenvector of $G^T$ belongs to the null space of $H_G$, since it has eigenvalue $1$. On the other hand, if $x\in{\rm Ker} (H_G)$, then $0=x^T H_G x = ((I-G)^T x)^T (I-G)^T x = \|(I-G)^T x\|_2^2$, which implies $(I-G)^T x = 0$, i.e., $x$ is an eigenvector of $G$ with eigenvalue $1$; therefore it is the dominant eigenvector. This fact is used, e.g., by \cite{PRadiabatic} in order to redefine the PageRank method as a ground-state method suitable for a quantum adiabatic computation.

\subsection{Quantum centrality}

Much of the literature on quantum centrality methods relies on discrete-time quantum walks. One notable example is the (discrete) Quantum PageRank, here abbreviated as DQPR \cite{Paparo}. For a graph with $n$ nodes, these methods typically work in an $n^2$-dimensional state space, as required by the coined or Szegedy formulations of the DTQW, although it can be proved that in DQPR the non-trivial component of the dynamics actually takes place in a subspace of dimension at most $2n$.

However, continuous-time quantum walks \cite{Feynman, FarhiGutmann, Childs} have also been used to define notions of graph centrality, mostly for undirected networks \cite{Izaac}. A CTQW on a graph ${\mathcal G}$ with $n$ vertices takes place in a complex Hilbert space of dimension $n$; a generic state of the system takes the form
$$
|\psi\rangle=\sum_{j=1}^n a_j |j\rangle,
$$ 
where $|j\rangle$, with $j=1,\ldots,n$, is the basis state associated with a walker localized at node $j$, and
 the amplitudes $a_1,\ldots,a_n\in\mathbb{C}$ are such that $\sum_{j=1}^n |a_j|^2=1$. Note that it holds $a_j=\langle j | \psi\rangle$. The square modulus $|a_j|^2$ is the {\em occupation} of node $j$, i.e., the probability of finding the system in the state $|j\rangle$ after a measurement. 
 
 The time evolution of a CTQW on a graph is described by the Schr\"odinger equation
\begin{equation}
i\frac{\partial |\psi(t)\rangle}{\partial t} = H |\psi(t)\rangle, \label{eq:schroedinger}
\end{equation}
where $|\psi(t)\rangle$ is the state of the system at time $t$. The Hamiltonian operator $H$ encodes the structure of the graph and is often chosen as the graph Laplacian or the adjacency matrix, which are symmetric when ${\mathcal G}$ is undirected; see \cite{Wong} for motivation and a comparison. 

If $H$ does not depend explicitly on time, the solution of \eqref{eq:schroedinger} is 
\begin{equation}
|\psi(t)\rangle=U(t) |\psi(0)\rangle,\label{eq:solution} \hspace{30pt}U(t)={\rm e}^{-iHt}.
\end{equation}
Here $U(t)$ is the unitary evolution operator of the system and $|\psi(0)\rangle$ is the initial state. Quantum properties of the CTQW include superposition of states and time reversibility; the latter is a consequence of the unitarity of the evolution operator and implies that there is no limit state. 

For this reason, one typically resorts to limits of time averages ({\em limiting distribution}) in order to define centrality and communicability scores. For instance, in \cite{Izaac} the authors define a CTQW-based centrality measure for undirected graphs where the initial state $|\psi(0)\rangle$ is a uniform superposition of all vertex states, the evolution of the walk is modeled by equation \eqref{eq:schroedinger} with $H=A$, and the centrality score of node $j$ is
\begin{equation}
C_j=\lim_{T\rightarrow\infty}\frac{1}{T}\int_0^T|\langle j|\psi(t)\rangle|^2 dt.\label{eq:Izaac}
\end{equation}

The following result, recalled from \cite{BoitoGrena} and adapted from \cite{Aharonov2}, ensures the well-posedness of the above definition and gives an explicit, closed-form characterization of the time-average limit.

\begin{theorem}\label{th:Thm1}
Let $U(t)={\rm e}^{-iHt}$ be a unitary evolution operator in a $n$-dimensional Hilbert space, and denote as $\{\theta_j, |\phi_j\rangle\}_{j=1,\ldots,n}$, the eigenvalues and eigenstates of the time-independent Hamiltonian $H$. Recall that the eigenvalues and eigenstates of $U(t)$ are $\{\lambda_j(t)= {\rm e}^{-i t\theta_j}, |\phi_j\rangle\}_{j=1,\ldots,n}$. Let $|\psi(0)\rangle=\sum_{i=1}^n a_j|\phi_j\rangle$, with $a_1,\ldots,a_n\in\mathbb{C}$, be the initial state of the system, written in the eigenstate basis, and denote as $|\psi(t)\rangle$ the state of the system at time $t$. Then for any state $|\xi\rangle$ it holds
\begin{equation}
\lim_{T\rightarrow\infty}\frac{1}{T}\int_0^T|\langle \xi|\psi(t)\rangle|^2 dt=
\sum_{j,k\,{\rm with }\, \theta_j=\theta_k }a_j a_k^* \langle\xi |\phi_j\rangle \langle\phi_k |\xi\rangle,\label{eq:prob}
\end{equation}
where the asterisk $^*$ denotes the complex conjugate.
\end{theorem}

Note that the limit depends on the initial state and on the eigenstates of $H$, but not on the eigenvalues of $H$, except for their multiplicity structure.

\section{Centrality for directed networks based on unitary CTQW}\label{sec:main}

It was mentioned in Section \ref{subs:classical} that both HITS and PageRank for directed networks can be seen as eigenvector problems on certain symmetric matrices. This suggests that we can define unitary CTQW-based centrality scores derived from classical HITS and PageRank, by using suitable modifications of these symmetric matrices as Hamiltonians.

The new methods proposed here follow a similar general scheme as in \cite{Izaac}:
\begin{enumerate}
\item Prepare the initial state (the ``walker'') in a state $|\psi(0)\rangle$.
\item Propagate the walker for a sufficiently large time $t$:
\begin{equation}
|\psi(t)\rangle={\rm e}^{-iHt}|\psi(0)\rangle.\label{eq:evolution}
\end{equation}
\item Compute the time-average probability distribution of finding the walker at each vertex $j$; this is the centrality score of node $j$:
\begin{equation}
C_j=\lim_{T\rightarrow\infty}\frac{1}{T}\int_0^T|\langle j|\psi(t)\rangle|^2 dt.\label{eq:centrality}
\end{equation}
\end{enumerate}
These centrality measures are well-defined, as a consequence of Theorem \ref{th:Thm1}.
Differently from \cite{Izaac}, however, $H$ here is not the adjacency matrix; moreover, the initial state can be different from the equal superposition of all vertex states.

\subsection{HITS--derived CTQW}

HITS looks for the dominant eigenvector of $A^TA$ (authority) or $AA^T$ (hubs). It is therefore quite natural to investigate the use of such matrices as Hamiltonians to define evolution operators for hub and authority rankings. We will describe the method for authority rankings; for hub rankings, simply exchange $A$ and $A^T$.

In fact, some difficulties arise if the matrix $A^TA$ is chosen as a Hamiltonian for authority rankings. Indeed, 
 $A^TA$ might have a block-diagonal structure; for instance, it is diagonal for a path graph. In this case, the resulting CTQW will be ``trapped'' in separate subspaces. This means that the quantum walk cannot explore the whole space and the resulting rakings might not be meaningful. In the example of a path graph, with a diagonal Hamiltonian, there is no evolution at all of the occupation of the nodes, because the states $|j\rangle$ are eigenvectors of the Hamiltonian, and the resulting ranking vector will be equal to the initial state. This is clearly not a desirable behaviour. For this reason we borrow from PageRank the idea of adding a rank-one correction to the adjacency matrix:  
\begin{equation}
\tilde{A}=\alpha A+\frac{1-\alpha}{n}{\mathbb{1}}{\mathbb{1}}^T,\label{eq:rank1}
\end{equation}
where $\alpha\in [0,1]$ is a suitably chosen parameter. Now define the following $n\times n$ symmetric matrix:
\begin{equation}
H_H= \tilde{A}^T \tilde{A}.\label{eq:CQHITShamiltonian}
\end{equation}
The matrix $H_H$ represents our Hamiltonian operator, which will be used to model a CTQW in a Hilbert space of dimension $n$ according to (\ref{eq:evolution}). In our experiments we take $\alpha=0.85$ as is often done for classical PageRank.

\subsection{PageRank--derived CTQW}

Direct PageRank (for authorities) seeks a vector in the 1-dimensional null space of $H_G=(I-G)(I-G)^T$, where $G$ is defined as in Section \ref{subs:classical}. As above, we will describe our PageRank-like method for authorities, and hubs can be found replacing $G$ with $G_h$.

Matrix $G$ already contains corrections that ensure the irreducibility of $H_G$; so, no further corrections are introduced. The Hamiltonian is
\begin{equation}
H_G = (I-G)(I-G)^T.\label{eq:CQPRhamiltonian}
\end{equation}
The Hamiltonian $H_G$ will be used to model a CTQW in a Hilbert space of dimension $n$ according to (\ref{eq:evolution}). The parameter $\alpha$ that defines the matrix $G$ is set to 0.85.

\subsection{Initial state}
As already mentioned, the choice of the initial state is required to properly define a quantum walk and its limiting distribution. In \cite{BoitoGrena} it was shown that the same evolution operator may give remarkably different rankings, depending on the choice of the initial state. 
Two options have been considered here for the initial state. The first one is simply a uniformly occupied state:
\begin{equation}
|\psi_u(0)\rangle=\frac{1}{\sqrt{n}}\sum_{k=1}^{n} |k\rangle. \label{eq:initialuniform}
\end{equation}

The second choice is an initial state where each node has occupation proportional to its in-degree (for authority centrality) or to its out-degree (for hub centrality):
\begin{eqnarray}
|\psi_w(0)\rangle=\frac{1}{\sqrt{\sum_{k=1}^{n} {\rm deg}_{\rm in}(k)}}\sum_{k=1}^{n} \sqrt{{\rm deg}_{\rm in}(k)}\,|k\rangle\hspace{10pt}\textrm{for authority},\nonumber\\
|\psi_w(0)\rangle=\frac{1}{\sqrt{\sum_{k=1}^{n} {\rm deg}_{\rm out}(k)}}\sum_{k=1}^{n} \sqrt{{\rm deg}_{\rm out}(k)}\,|k\rangle\hspace{10pt}\textrm{for hub}.\label{eq:initialweighted}
\end{eqnarray}
This choice of initial state is mainly motivated by the fact that $|\psi_w(0)\rangle$ is at the same time easy to compute and strongly correlated with the HITS score vector. Numerical experiments \cite{BoitoGrena} also confirm its effectiveness.

Combining the two Hamiltonians defined above with these two choices of the initial state, we obtain four quantum centrality methods:
\begin{itemize}
\item{} CQHITSu: Hamiltonian $H_H$ (\ref{eq:CQHITShamiltonian}), initial state $|\psi_u(0)\rangle$ (\ref{eq:initialuniform});
\item{} CQHITSw: Hamiltonian $H_H$ (\ref{eq:CQHITShamiltonian}), initial state $|\psi_w(0)\rangle$ (\ref{eq:initialweighted});
\item{} CQPRu: Hamiltonian $H_G$ (\ref{eq:CQPRhamiltonian}), initial state $|\psi_u(0)\rangle$ (\ref{eq:initialuniform});
\item{} CQPRw: Hamiltonian $H_G$ (\ref{eq:CQPRhamiltonian}), initial state $|\psi_w(0)\rangle$ (\ref{eq:initialweighted}).
\end{itemize}

\section{Numerical experiments}\label{sec:experiments}

The four proposed methods have been tested on small graphs, where clear hub / authority results can be expected, in order to spot possible drawbacks of the methods, and then tested on a large sample of larger graphs generated using the Python package NetworkX \cite{NetworkX}. All the methods were implemented in Octave and run on a laptop equipped with a 4-core Intel i7-7500U 2.70GHz processor, with 16 GB of RAM. In this paper we do not attempt a quantum implementation of the proposed centrality measures: all the numerical tests involve classical computations. The limit in \eqref{eq:centrality} is computed via Theorem \ref{th:Thm1}.

This section summarizes the main results obtained for the various methods.

\subsection{Small graphs}

The small graphs chosen for the initial test are:

\begin{description}
\item{Example 1:} path graph, i.e., a simple chain of nodes.
\item{Example 2:} diamond graph, i.e., a graph with a ``main hub'' (node 1) with $n-2$ outgoing edges directed towards nodes $2,3,...,n-1$, and a ``main authority'' (node $n$) with $n-2$ incoming edges from nodes $2,3,...n-1$.
\item{Example 3:} star graph, i.e., a central hub (node 1) connected to all the other nodes, and no other edge present.
\item{Example 4:} tailed graph, i.e., $n_1$ initial nodes connected as in a path graph, followed by $n_2$ nodes that form a complete subgraph. Node $n_1$ is connected to all the $n_2$ nodes in the complete subgraph.
\end{description}

\begin{figure}
\centering
\includegraphics[width=0.6\textwidth]{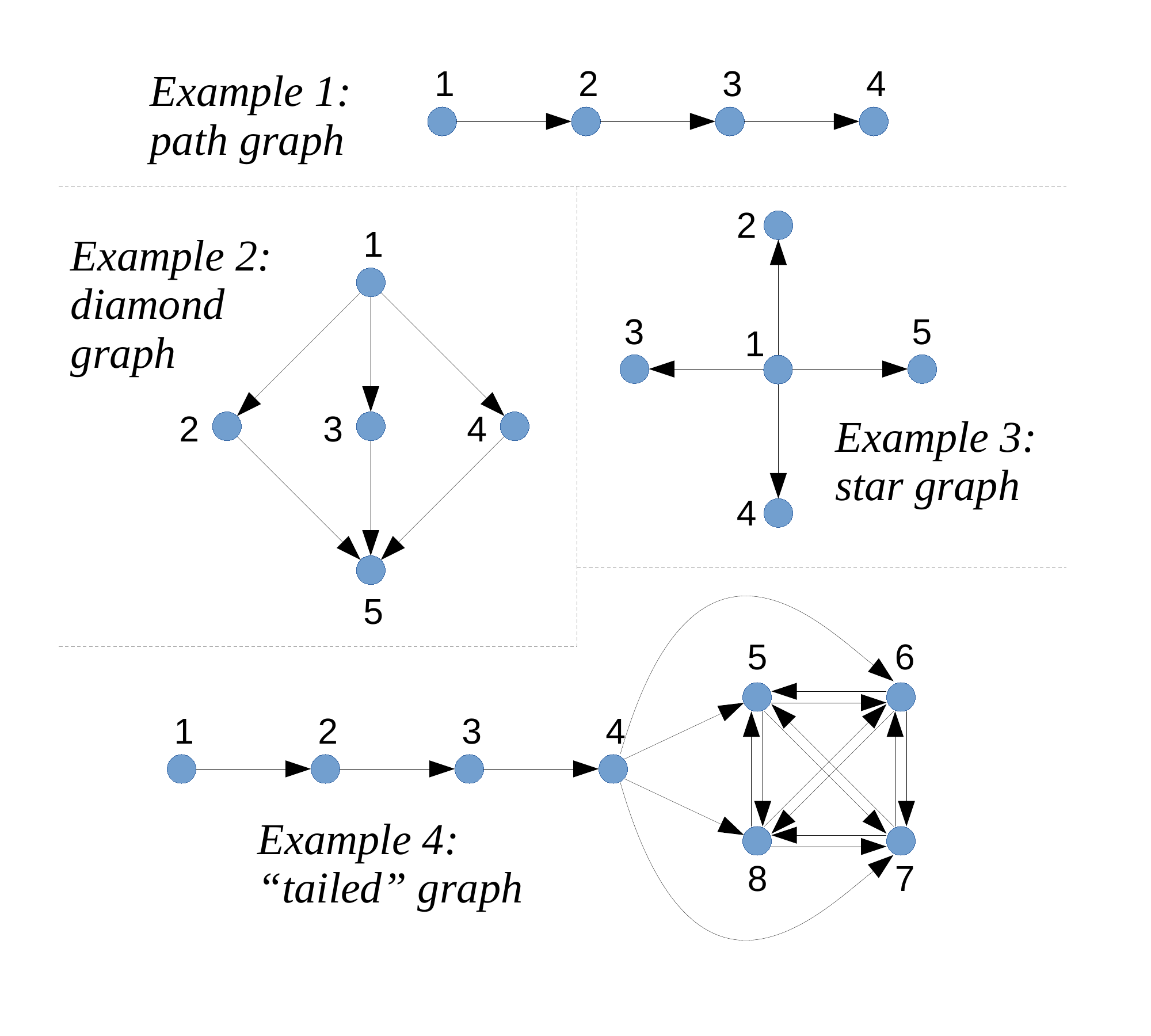}
\caption{Simple graphs used for testing the ranking methods (examples 1-4).}\label{fig:simplegraphs}
\end{figure}

These graphs are shown in Figure \ref{fig:simplegraphs}. The computed centrality scores for Examples 1--3 and the rankings for Example 4 are shown in 
Tables 1--6. The rankings provided by the quantum methods are typically similar to the rankings given by their classical counterparts. However, some details are worth discussing.

\begin{table}[!h]
\begin{center}
\caption{Hub scores for Example 1, with 4 nodes. Authority scores can be read by inverting the order of the nodes.}\label{table:expath}
\begin{tabular}{l l l l l l l}
\hline
Node & CQHITSu & CQHITSw & HITS & CQPRu & CQPRw & PR\\
\hline\\
1 & 0.2683 & 0.3301 & 0.5774 & 0.4541 & 0.4479 & 0.3701\\
2 & 0.2683 & 0.3301 & 0.5774 & 0.2795 & 0.3147 & 0.2988\\
3 & 0.2683 & 0.3301 & 0.5774 & 0.1820 & 0.1636 & 0.2149\\
4 & 0.1952 & 0.0097 & 0.0000 & 0.0844 & 0.0737 & 0.1161\\
\hline
\end{tabular}
\end{center}
\end{table}

\begin{table}[!h]
\begin{center}
\caption{Hub scores for Example 2, with 5 nodes. Authority scores can be read by exchanging node 1 and node 5.}\label{table:exdiamond}
\begin{tabular}{l l l l l l l}
\hline
Node & CQHITSu & CQHITSw & HITS & CQPRu & CQPRw & PR\\
\hline\\
1 & 0.4055 & 0.4886 & 0.5000 & 0.5606 & 0.6787 & 0.4683\\
2 & 0.1400 & 0.1695 & 0.5000 & 0.0955 & 0.0879 & 0.1407\\
3 & 0.1400 & 0.1695 & 0.5000 & 0.0955 & 0.0879 & 0.1407\\
4 & 0.1400 & 0.1695 & 0.5000 & 0.0955 & 0.0879 & 0.1407\\
5 & 0.1746 & 0.0028 & 0.0000 & 0.1528 & 0.0578 & 0.1096\\
\hline
\end{tabular}
\end{center}
\end{table}

\begin{table}
\begin{center}
\caption{Hub scores for Example 3, with 5 nodes.}\label{table:exstarhub}
\begin{tabular}{l l l l l l l}
\hline
Node & CQHITSu & CQHITSw & HITS & CQPRu & CQPRw & PR\\
\hline\\
1 & 0.2599 & 0.9906 & 1.00000 & 0.5685 & 0.7162 & 0.5238\\
2 & 0.1850 & 0.0023 & 0.00000 & 0.1079 & 0.0710 & 0.1190\\
3 & 0.1850 & 0.0023 & 0.00000 & 0.1079 & 0.0710 & 0.1190\\
4 & 0.1850 & 0.0023 & 0.00000 & 0.1079 & 0.0710 & 0.1190\\
5 & 0.1850 & 0.0023 & 0.00000 & 0.1079 & 0.0710 & 0.1190\\
\hline
\end{tabular}
\end{center}
\end{table}

\begin{table}
\begin{center}
\caption{Authority scores for Example 3, with 5 nodes.}\label{table:exstarauth}
\begin{tabular}{l l l l l l l}
\hline
Node & CQHITSu & CQHITSw & HITS & CQPRu & CQPRw & PR\\
\hline\\
1 & 0.1850 & 0.0007 & 0.0000 & 0.1491 & 0.2484 & 0.1709\\
2 & 0.2037 & 0.2498 & 0.5000 & 0.2127 & 0.1879 & 0.2073\\
3 & 0.2037 & 0.2498 & 0.5000 & 0.2127 & 0.1879 & 0.2073\\
4 & 0.2037 & 0.2498 & 0.5000 & 0.2127 & 0.1879 & 0.2073\\
5 & 0.2037 & 0.2498 & 0.5000 & 0.2127 & 0.1879 & 0.2073\\
\hline
\end{tabular}
\end{center}
\end{table}

\begin{table}
\begin{center}
\caption{Hub rankings for Example 4, with $n_1=n_2=4$.}\label{table:extailhub}
\begin{tabular}{l l l l l l l}
\hline
CQHITSu & 4 & 1,2,3   &   &   & 5,6,7,8 &       \\
CQHITSw & 4 & 5,6,7,8 &   &   &         & 1,2,3 \\
HITS    & 4 & 5,6,7,8 &   &   &         & 1,2,3 \\
CQPRu   & 1 & 2       & 3 & 4 & 5,6,7,8 &       \\ 
CQPRw   & 1 & 2       & 3 & 4 & 5,6,7,8 &       \\
PR      & 1 & 2       & 3 & 4 & 5,6,7,8 &       \\
\hline
\end{tabular}
\end{center}
\end{table}

\begin{table}
\begin{center}
\caption{Authority rankings for Example 4, with $n_1=n_2=4$.}\label{table:extailauth}
\begin{tabular}{l l l l l l}
\hline
CQHITSu & 5,6,7,8 & 2,3,4   &       &   & 1 \\
CQHITSw & 5,6,7,8 & 2,3,4   &       &   & 1 \\
HITS    & 5,6,7,8 & 1,2,3,4 &       &   &   \\
CQPRu   & 5,6,7,8 & 3       & 4     & 2 & 1 \\
CQPRw   & 5,6,7,8 & 4       & 3     & 2 & 1 \\
PR      & 5,6,7,8 & 4       & 3     & 2 & 1 \\
\hline
\end{tabular}
\end{center}
\end{table}

In \cite{BoitoGrena}, a ``quantum pathology'' affecting almost all the tested quantum methods (including DQPR) was spotted in small examples similar to these ones. When a graph contains ``no-hub'' nodes, i.e., nodes with zero out-degree, or ``no-authority'' nodes, any sensible ranking should place them at the bottom of the respective rankings. However, this is not always the case for quantum methods, and this problem sometimes also appears in the four centrality measures presented here. CQHITSu and CQPRu fail in Example 2: node 5, which has no outgoing edges, is placed second in hub rankings instead of being in the last position, and the same happens to node 1 for authority rankings. CQPRw fails in the authority ranking in Example 3: node 1, the central node of the star (with no incoming edges) is ranked first instead of last. No evident pathologies emerge from Examples 1 and 4.

The only method that seems to be immune to ``quantum pathologies'' is CQHITSw. In fact, the introduction of the initial weighted state was aimed at correcting this pathology; it is effective for HITS-derived methods, but the behaviour of PR-derived methods seems to be more erratic, since in Example 3 the pathology unexpectedly {\it appears} as an effect of the weighted initial vector. 

We add some more remarks:
\begin{itemize}
\item{} In path graphs, different behaviours are observed for HITS, which isolates the last node (hub) or the first node (authority) placing the other ex-aequo, and for PR, which outputs descending hub scores and ascending authority scores, without ties. Both behaviours are acceptable; it is interesting to see which one is reproduced by the new methods. It appears that the two HITS-derived methods have the same behaviour as HITS, and the two PR-derived methods have the same behaviour as PR, as expected.
\item{} Classical HITS shows a notable failure on Example 2 (it does not isolate the first-ranked node, both in hub and authority rankings) and a less significant failure in Example 4 (node 1 is given the same authority as nodes 2, 3 and 4). In Example 2, these problems are likely linked to the degeneration of the dominant eigenvector.  Note that the two CQHITS method do not suffer from these drawbacks and correctly spot the first-ranked node in Example 2 (and also the last-ranked node in Example 4), isolating them from the others.
\item{} An interesting feature of Example 4 is the fact that classical PR and HITS output different hub rankings: PR places node 1 first, followed by the other nodes of the ``tail'' -- thus emphasizing the fact that all the following nodes can be reached from them -- while HITS privileges node 4, emphasizing the number of direct connections. It is noteworthy that the quantum versions reproduce this different behaviour.
\item{} In Example 4, CQHITSu identifies the same top hub as HITS, but it places nodes 1-3 above nodes 5-8, contrarily to HITS (and CQHITSw); in the same example, there is a curious inversion of nodes 3 and 4 in the authority ranking for CQPRu.
\end{itemize}

\subsection{Tests on larger graphs}\label{sec:largegraphs}

\subsubsection{Scale-free graphs}\label{subsubsec:scalefree}
Here we test ranking algorithms on 2174 larger scale-free graphs authomatically generated by the {\tt scale\_free\_graph} command in NetworkX \cite{NetworkX}. These are directed graphs built according to a preferential attachment process. The number of graphs generated for each choice of the number of nodes is given in Table \ref{table:nnodes}.
\begin{table}
\begin{center}
\caption{Number of scale-free graphs used in numerical tests and corresponding number of nodes.}\label{table:nnodes}
\begin{tabular}{l | l l l l l l l l}
\hline
Number of nodes & $128$  & $256$ & $384$ & $512$ & $640$ & $768$ & $896$ & $1024$\\
Number of graphs & $800$  & $400$ & $267$  & $200$ & $160$ & $133$ & $114$ & $100$\\
\hline
\end{tabular}
\end{center}
\end{table}
 Tests are repeated for hub and authority rankings, so each graph is used twice.

Table \ref{table:scalefree} shows the comparison of the four centrality methods with their corresponding classical counterpart (HITS for CQHITSu and CQHITSw, PR for CQPRu and CQPRw), showing
\begin{itemize}
\item{} the fraction $F_1$ of tests in which the first node is the same as found in the corresponding classical algorithm,
\item{} the average number $F_{10}$ of the top 10 nodes that appear in both the quantum and classical rankings, and
\item{} the Kendall's $\tau$ \cite{Ktau} of the overall ranking, w.r.t.~the ranking obtained by the classical algorithm.
\end{itemize}

\begin{table}
%\begin{center}
\caption{$F_1$, $F_{10}$ and Kendall's $\tau$ for scale-free graphs.}\label{table:scalefree}
\hskip-2.0cm\begin{tabular}{l| l l l| l l l |l l l |l l l}
\hline
\multirow{2}{*}{Nodes}& \multicolumn{3}{c}{CQHITSu}&
\multicolumn{3}{c}{CQHITSw}&\multicolumn{3}{c}{CQPRu}&
\multicolumn{3}{c}{CQPRw}\\
& $F_1$ & $F_{10}$ & $\tau$ & $F_1$ & $F_{10}$ & $\tau$ & $F_1$ & $F_{10}$ & $\tau$ & $F_1$ & $F_{10}$ & $\tau$ \\
\hline 
128  & 0.841 & 6.22 & 0.0288 & 0.900 & 7.68 & 0.7000 & 0.917 & 8.27 & 0.3198 & 0.978 & 9.23 & 0.5658 \\
256  & 0.804 & 6.69 & 0.0026 & 0.893 & 8.09 & 0.6940 & 0.933 & 8.29 & 0.2652 & 0.984 & 9.19 & 0.5225 \\
384  & 0.768 & 6.54 & 0.0019 & 0.869 & 7.85 & 0.6697 & 0.901 & 8.26 & 0.2718 & 0.963 & 9.15 & 0.5118 \\
512  & 0.758 & 6.67 & -0.0205 & 0.855 & 7.87 & 0.6754 & 0.910 & 8.23 & 0.2321 & 0.970 & 9.16 & 0.4894 \\
640  & 0.731 & 6.68 & -0.0088 & 0.872 & 7.87 & 0.6671 & 0.922 & 8.07 & 0.2363 & 0.972 & 9.05 & 0.4850 \\
768  & 0.695 & 6.42 & -0.0024 & 0.831 & 7.56 & 0.6508 & 0.914 & 7.95 & 0.2520 & 0.970 & 8.96 & 0.4878 \\
896  & 0.667 & 6.55 & -0.0123 & 0.851 & 7.80 & 0.6523 & 0.882 & 7.89 & 0.2406 & 0.961 & 8.95 & 0.4750 \\
1024 & 0.685 & 6.51 & -0.0088 & 0.865 & 7.87 & 0.6545 & 0.920 & 7.78 & 0.2296 & 0.975 & 8.91 & 0.4705 \\
\hline
\end{tabular}
%\end{center}
\end{table}

We see that HITS-derived algorithms perform quite well, with a significant improvement of CQHITSw over CQHITSu. CQHITSw finds the same first node as HITS in 83--90\% of the tests, with an average coincidence of more than 7 nodes in the top 10. Results for CQHITSu are still acceptable but clearly worse. Note the marked difference in Kendall's $\tau$ parameter for the two methods: about $0$ for CQHITSu, and above $0.65$ for CQHITSw. The $\tau$ value for CQHITSu indicates that, although the method is generally able to find top-ranked nodes, the overall rankings are weakly related to HITS rankings. This is almost certainly due to the above-mentioned quantum pathology, which disrupts rankings in lower positions; see \cite{BoitoGrena} for an extensive discussion. CQHITSw, on the contrary, exhibits the highest values of $\tau$ among the four methods.

PR-derived methods perform even better than HITS-derived methods in finding top-ranked nodes: these are correctly identified in about 90\% of the cases for CQPRu and in more than 96\% of the cases for CQPRw. The number of ``correct'' nodes among the top ten is also remarkable: about 8 for CQPRu and about 9 for CQPRw. Kendall's $\tau$, however, is not high (usually below 0.3 for CQPRu and around 0.5 for CQPRw). The improvement achieved adopting a weighted initial vector is significant, but it is less marked than for HITS-derived methods. A possible explanation might come from the fact that the HITS ranking is known to be strongly correlated to the simple degree ranking, while the correlation is weaker for PR. 

A regular trend can be spotted in the $F_1$ parameter for CQHITSu, as it appears to decrease when increasing the size of the graphs. Such a behaviour is not observed in other methods. 

\subsubsection{Random $k$-out graphs}

Another test was performed on random graphs built via preferential attachment, with the constraint that all nodes have (nearly) the same outdegree. These graphs were generated by the {\tt random\_k\_out\_graph} command in NetworkX \cite{NetworkX} and then modified by removing multiple edges, which is the reason for the slight deviation from uniform outdegree.
 A sample of 3000 graphs of 128 nodes each was considered. The same quantities $F_1$, $F_{10}$ and $\tau$ are computed for the four methods, as in \ref{subsubsec:scalefree}. Since the roles of hubs and authorities are asymmetrical due to the outdegree constraint, separate averages are computed. Results are shown in Table \ref{table:random}.

\begin{table}
%\begin{center}
\caption{$F_1$, $F_{10}$ and Kendall's $\tau$ for random graphs.}\label{table:random}
\hskip-2.0cm\begin{tabular}{l| l l l| l l l |l l l |l l l}
\hline
\multirow{2}{*}{}& \multicolumn{3}{c}{CQHITSu}&
\multicolumn{3}{c}{CQHITSw}&\multicolumn{3}{c}{CQPRu}&
\multicolumn{3}{c}{CQPRw}\\
& $F_1$ & $F_{10}$ & $\tau$ & $F_1$ & $F_{10}$ & $\tau$ & $F_1$ & $F_{10}$ & $\tau$ & $F_1$ & $F_{10}$ & $\tau$ \\
\hline
Hub   & 0.923 & 9.43 & 0.8877 & 0.923 & 9.42 & 0.8854 & 0.958 & 9.59 & 0.8955 & 0.959 & 9.60 & 0.9007 \\
Auth. & 0.972 & 7.19 & -0.0640 & 0.991 & 9.12 & 0.8136 & 0.785 & 7.67 & 0.4174 & 0.988 & 9.72 & 0.7944 \\
\hline
\end{tabular}
%\end{center}
\end{table}

For hub rankings, results for weighted methods are almost identical to results for a uniform initial vector, as can be expected. The correspondence with classical methods seems to be very good, even -- quite unexpectedly -- for Kendall's $\tau$ (around 0.9). In authority ranking, uniform methods give good results -- except for Kendall's $\tau$, as expected -- but the improvement from uniform to weighted methods is significant: the weighted methods show a striking 99\% probability of spotting the same first-ranked node as their classical counterparts, an average coincidence of more than 9 among the first 10 nodes and a Kendall's $\tau$ around 0.8.

\section{Discussion}\label{sec:discussion}
One remarkable feature of the methods presented here is their overall good (sometimes  {\em very} good) accordance with classical methods such as HITS or PageRank. One may ask why, since the classical methods seek the eigenvector associated with a specific eigenvalue (the dominant or the null eigenvalue), whereas from Theorem \ref{th:Thm1} it is clear that eigenvalues have no role at all in the quantum rankings, except possibly for their multiplicity structure. Indeed, suppose for simplicity that the Hamiltonian matrix $H$ has distinct eigenvalues: then the limiting distribution of the associated CTQW only depends on the eigenvectors of $H$ and on the initial state. Why should CTQW-based centrality give similar results to the classical case, if CTQWs do not ``see'' eigenvalues? 
%The same question obviously applies to discrete-time quantum centrality, although here we focus on the continuous-time case.

\begin{figure}
\begin{center}
\caption{Coefficients $|a_j|^2$ for the CQPRu authority ranking of a scale-free graph with 128 nodes and parameters $\alpha=0.557$, $\beta=0.136$, $\gamma=0.307$. Here $|a_1|^2=0.5078$, whereas the other coefficients are in the $[0,0.0608]$ range.}\label{fig:acoeff1}
\includegraphics[width=\textwidth]{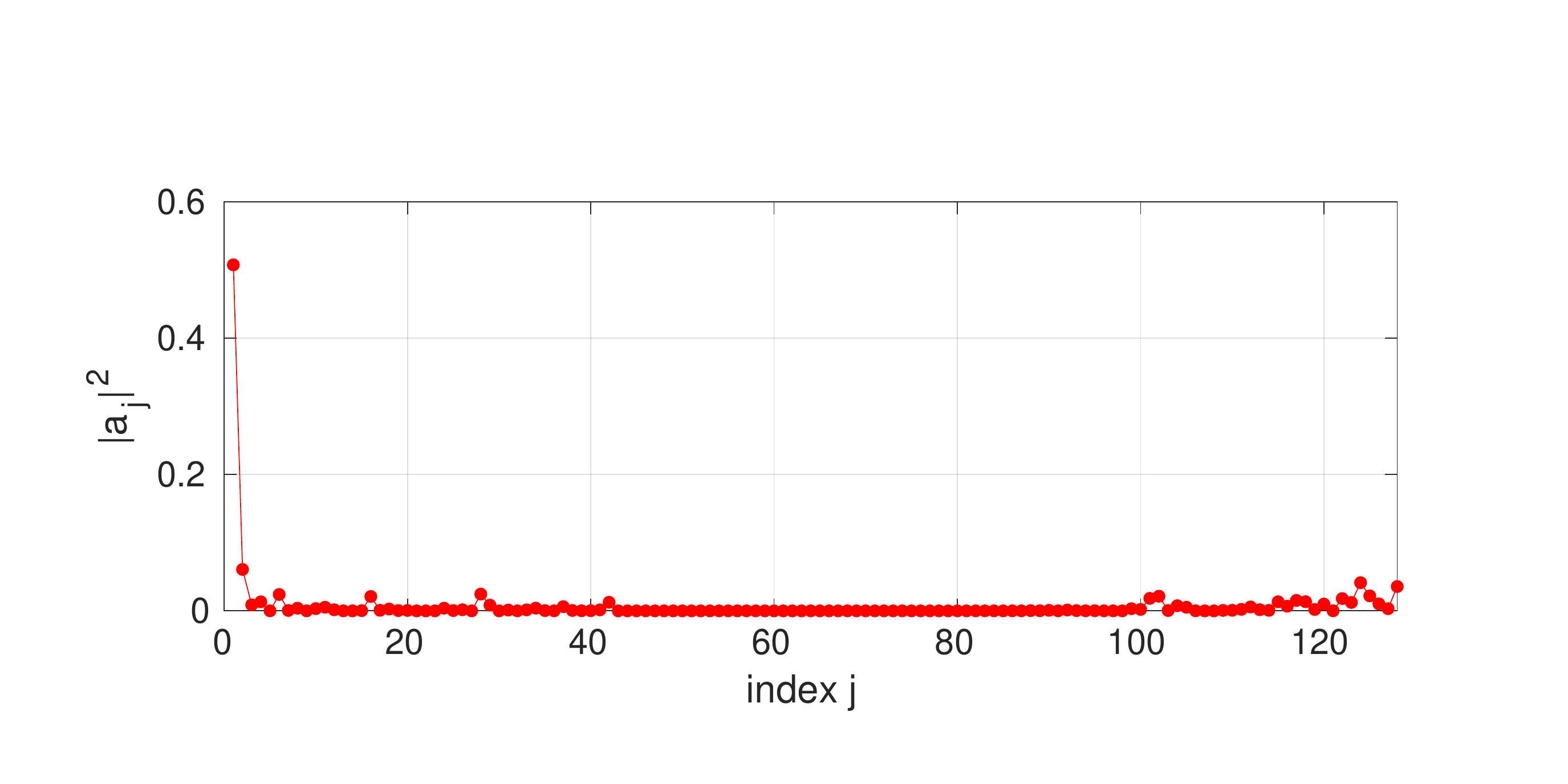}
\end{center}
\end{figure}

\begin{figure}
\begin{center}
\caption{Coefficients $|a_j|^2$ for the CQPRu hub ranking of a $k$-out graph with 128 nodes and parameters $\alpha=0.3$, $k=5$. Here $|a_1|^2=0.9249$, whereas the other coefficients are in the $[0,0.0077]$ range.}\label{fig:acoeff2}
\includegraphics[width=\textwidth]{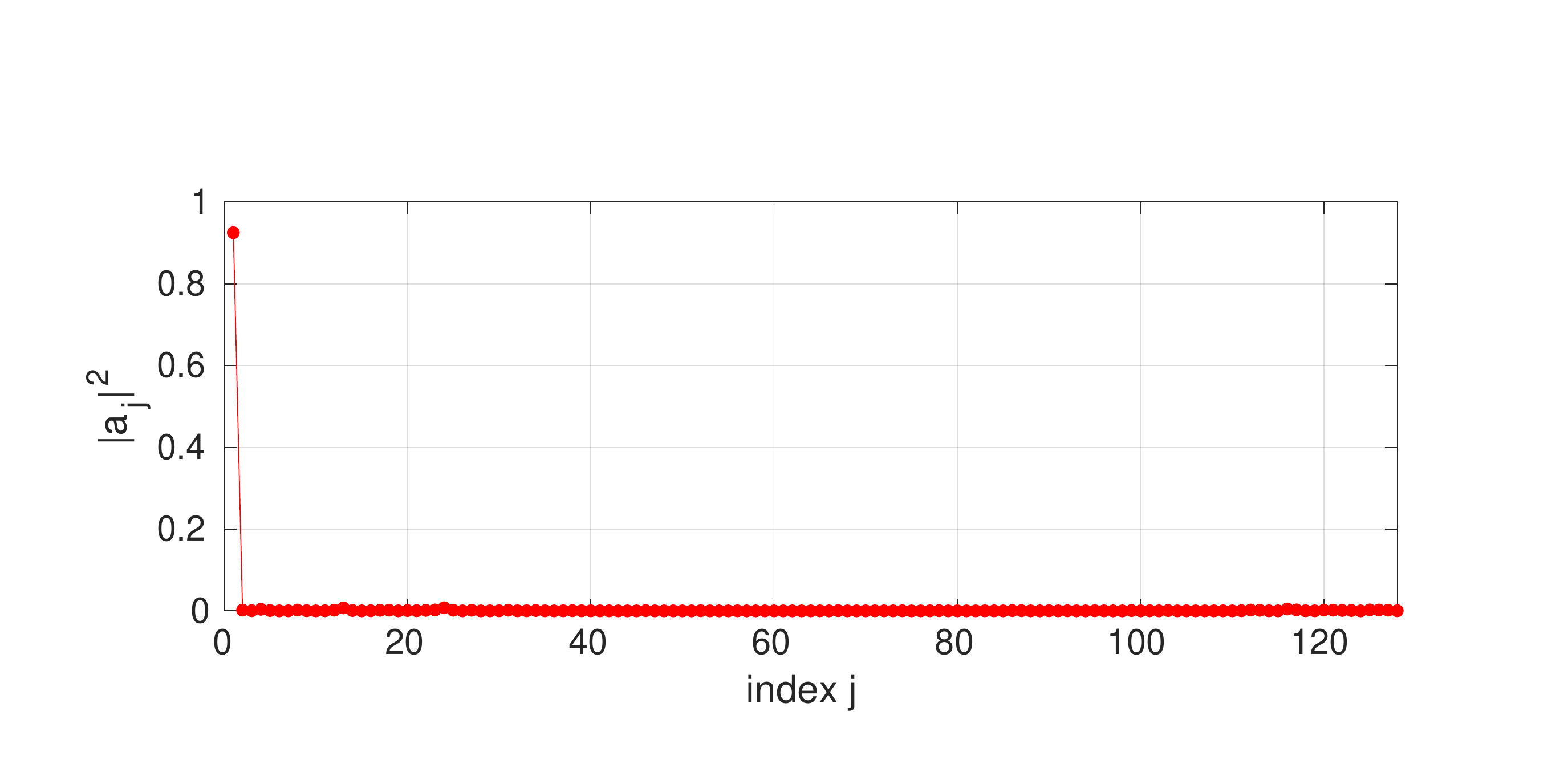}
\end{center}
\end{figure}

The explanation we found to be most convincing is the following. The classical ranking vector, say $\phi_1$, is positive by the Perron-Frobenius theorem. Now, recall that coefficient $a_1$ from Theorem \ref{th:Thm1} is the (possibly weighted) sum of the elements of $\phi_1$. Then one can typically expect that $|a_1|\gg |a_j|$ for $j>1$, thus amplifying the contribution of eigenvector $\phi_1$ in the quantum ranking. Figures \ref{fig:acoeff1} and \ref{fig:acoeff2} show the behaviour of the coefficients $|a_j|^2$ for a scale-free and for a $k$-out graph, respectively, for the CQPRu method. In both cases $a_1$ is the coefficient associated with the classical PageRank eigenvector and it is clearly larger than the others. Note that both rankings are quite faithful to their classical counterparts: they identify the same top-ranked node and the same set of top-ten nodes.

\begin{figure}
\begin{center}
\caption{Entries of a few eigenvectors of the Hamiltonian matrix for the CQPRu hub ranking of a $k$-out graph with 128 nodes and parameters $\alpha=0.3$, $k=5$. The first eigenvector (thick blue line) is the classical PageRank vector, i.e., it is associated with eigenvalue $0$. Also depicted are the eigenvectors associated with eigenvalues $0.2908$, $0.9517$ and $1.4143$, that is, columns of indices $3$, $50$ and $100$ in the eigenvector matrix computed by MATLAB. Clearly the first eigenvector has a comparatively uniform behaviour, whereas the other eigenvectors are much more localized.}\label{fig:eigenvectors}
\includegraphics[width=0.8\textwidth]{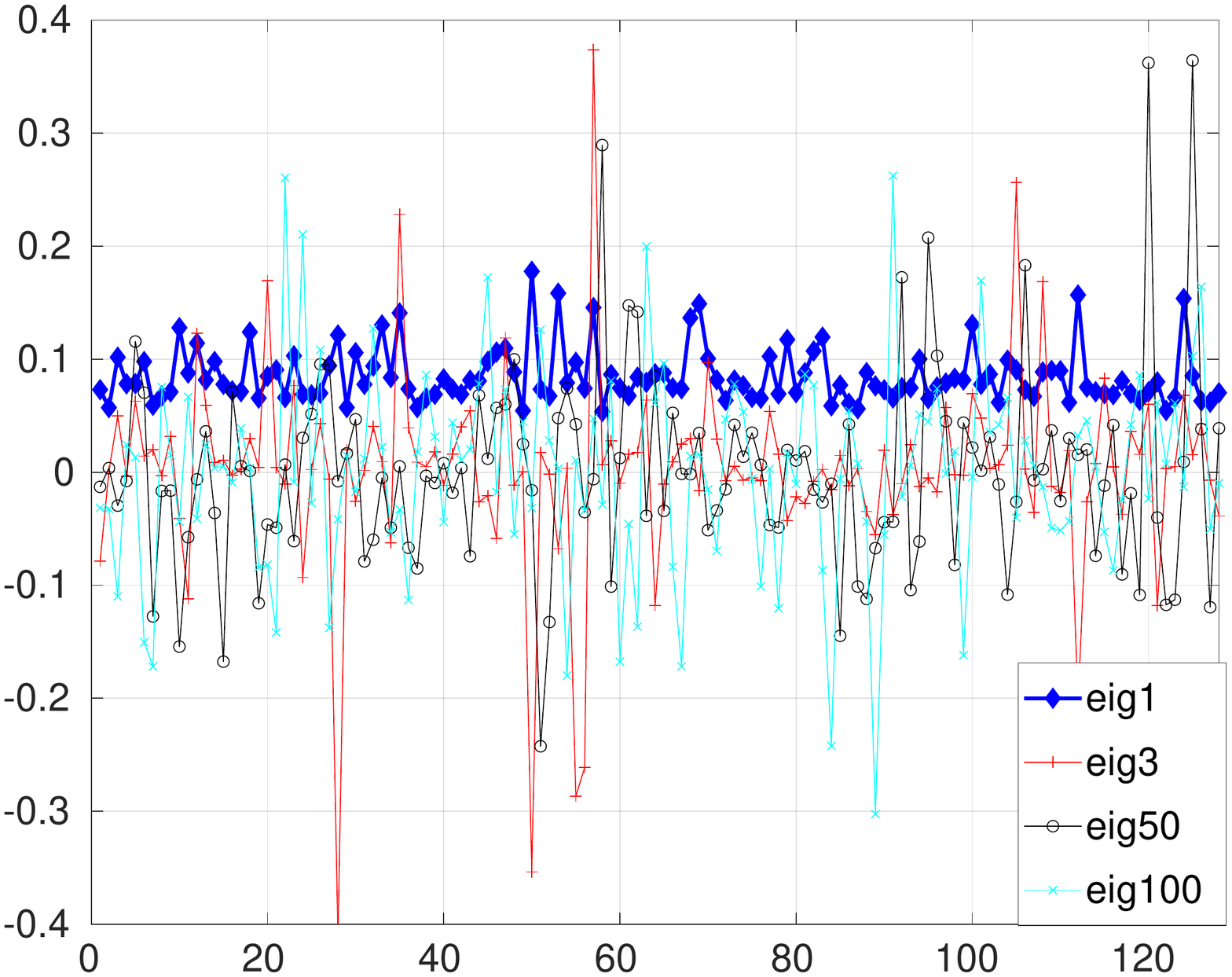}
\end{center}
\end{figure}

This remark may also help explain why hub rankings for $k$-out graphs are in extremely good accordance with classical PageRank: since the PageRank vector is close to being uniform (as shown in Figure \ref{fig:eigenvectors}), the value of $a_1$ is nearly maximised, so in Theorem \ref{th:Thm1} the term associated with the PageRank eigenvector is strongly dominant.

One may also ask whether the behaviour of the quantum methods is related to localization properties of eigenvectors. One may conjecture, for instance, that quantum approaches amplify the contribution of localized eigenvectors. Localization of eigenvector centrality is a well-known phenomenon: see, e,g., \cite{Goh01, Martin14, Metz21}. However, in our experiments the classical ranking eigenvector does not appear to be remarkably more localized than others. Moreover, the various eigenvectors tend to be localized over different sets of indices, so they do not necessarily reinforce the contribution of the classical ranking eigenvector. Hub rankings for $k$-out graphs are a good counterexample for this conjecture: the classical ranking eigenvector is typically much less localized than other eigenvectors (see Figure \ref{fig:eigenvectors}), but, as noted above, the quantum rankings are in very good accordance with classical rankings.

Another notable phenomenon is the fact that, for the CQPRu method, many coefficients $a_i$ may be zero, and therefore the associated eigenvectors of the Hamiltonian do not contribute to the ranking. For instance, in the example of Figure \ref{fig:acoeff1}, $a_j$ is numerically zero for $j=21, 22, 43$ to $80, 98, 107, 108$. % We do not see this phenomenon in the k-ou graph example.
More precisely, the Hamiltonian has many eigenvectors with eigenvalue 1 that are orthogonal to the initial uniform state. In fact, in such cases
$G$ and $G^T$ both have large-dimensional null spaces, so it is not surprising that the intersection of these null spaces also has a comparatively large dimension. Now ${\rm Ker}(G)\cap {\rm Ker}(G^T)$ is an eigenspace of $H$ associated with the eigenvalue $1$, and each vector in this subspace has zero average, i.e., it is orthogonal to the uniform vector. Indeed, let ${\bf e}$ be the vector with all entries equal to 1. Since $G$ is row-stochastic we have ${\bf e}^TG^T={\bf e}^T$. Then for any ${\bf v}\in{\rm Ker}(G)\cap {\rm Ker}(G^T)$ it holds ${\bf e}^TG^T{\bf v}={\bf e}^T{\bf v}$. But $G^T{\bf v}=0$ and therefore ${\bf e}^T{\bf v}=0$. 

\section{Conclusions}\label{sec:conclusions}
We tested node-ranking methods based on continuous-time quantum walks, obtained by reframing two well-known classical methods (HITS and PageRank) as eigenvector problems on symmetric matrices, and using such matrices (with a slight modification for HITS) as Hamiltonians for the quantum walk. Each Hamiltonian was tested on two initial vectors (uniform and degree-weighted), yielding four quantum ranking methods. Numerical results are in good agreement with their classical counterparts in finding the top-ranked nodes, with a significant improvement in weighted methods w.r.t.~the uniform ones. For instance, CQPRw finds the same top-ranked node as classical PageRank in more than 95\% of the cases, and the average number of common top-10 nodes is always above 8.9, for all sets of test graphs. CQHITSw seems to be immune to the anomalous rankings of no-authority or no-hub nodes that are frequently encountered when using quantum methods, and has the best performance for the overall agreement of the rankings, as measured by Kendall's $\tau$. Analysis of the results shows that the reason for this agreement is not related to localization of the dominant (or null) eigenvector found in the classical case. It is more likely due to the Perron-Frobenius theorem, which ensures that the initial positive state (uniform or weighted) has a large component along the eigenvector corresponding to the classical ranking vector.

\section*{Acknowledgment}

This work was partially supported by the project PRA\_2020\_92 ``Quantum computing, technologies and application'' funded by the University of Pisa.

\end{document}